\documentclass{article}
\usepackage{spconf,amsmath,graphicx,amssymb}
\usepackage{algorithm}
\usepackage{algorithmic}

\title{Vertex-disjoint Cycle Cover for graph signal processing }
\name{Raghavendra Singh}
\address{IBM Research India\\raghavsi@in.ibm.com}
\begin{document}
%
\maketitle
\begin{abstract}
Eigenvectors of the Laplacian of a cycle graph  exhibit the sinusoidal characteristics of the standard DFT basis, and signals defined on such graphs are amenable to linear shift invariant (LSI) operations. In this paper we propose to reduce a generic graph to its vertex-disjoint cycle cover, i.e.,  a set of  subgraphs that are cycles, that together contain all vertices of the graph, and no two subgraphs have any vertices in common. 
Additionally if the weight of an edge in the graph is a function of the variation in the signals on its vertices, then maximally smooth cycles can be found, such that the resulting DFT does not have high frequency components. We show that an image graph  can be reduced to such low-frequency cycles, and use that to propose a simple image denoising algorithm.

\end{abstract}
\section{Introduction}
\label{sec:intro}
In this interconnected world it is recognised that signals could lie on the vertices of large graphs, such as, user's preferences in social networks, or sensing data in sensor networks.  Edges between the vertices define the dependencies between the signal on the vertices, for example ``neighbouring" pixels in an image take on similar values. Recently there has been considerable interest in processing signals on graphs, see e.g.,~\cite{ShuNarFroOrtVan,CouGraNajPesTal}. Challenges include defining neighbourhoods for an arbitrary topology, and processing dependencies along arbitrary paths~\cite{ShuNarFroOrtVan}. 

An approach towards  tackling these challenges exploits the spectral properties of the Laplacian matrix of the graph~\cite{NarOrt,HamVanGri,AgaLu,EkaFanAyaRam}. A Graph Fourier Transform (GFT) that extends Fourier analysis to signals  on a graph can be defined using the eigenvectors of the graph Laplacian. This approach is intuitively and mathematically justified - Laplacian of a graph is equivalent to the Laplace operator in time domain, eigenfunctions of the latter are the DFT basis, hence eigenvalues of the former can be considered to be ``GFT" basis that furthermore capture a notion of smoothness of the signal with respect to the graph topology~\cite{ShuNarFroOrtVan}. 

A class of graphs for which DFT and GFT are equivalent (upto a permutation) are circulant graphs~\cite{AgaLu, EkaFanAyaRam, GraPol}. As expected (and desired) signals on circulant graphs are shift invariant much like discrete time periodic signals. Thus these graphs, unlike generic graphs, are amenable to liner shift invariant processing -- in \cite{EkaFanAyaRam} the authors have shown that  operations such as shifting and sampling are natural to these graphs, and that non-circulant graphs can be decomposed into circulant graphs, similar to a linear time variant system being represented as a bank of linear time invariant systems.

Cycle graphs are a sub-class of circulant graphs; they are 2-regular, hence the sparsest possible circulant graphs. In graph theory there is considerable significance attached to cycles in a graph, e.g., Eulerian, Hamiltonian cycles, the cycle basis of a graph and so on. We are interested in a vertex cycle cover of a graph which is a set of cycles that are subgraphs,  and contain all vertices of the graph. If the cycles of the cover have no vertices in common, the cover is called vertex-disjoint cycle cover. In this case the set of the cycles constitutes a spanning subgraph of the graph. A vertex-disjoint cycle cover of an undirected graph (if it exists) can be found in polynomial time by transforming the problem into a problem of finding a perfect matching in a larger graph~\cite{Tut}. If weights are defined on the edges then minimum weight cycle cover is equivalent to minimum weight perfect matching which can be found in polynomial time using Edmonds algorithm~\cite{Edm}. Note that these weighted cycles do not necessarily have circulant Laplacians (these matrices are Hankel not necessarily Toeplitz), hence their eigenvector space may not be DFT like. However if only the structure of the cycle graph is preserved and weights are ignored (or small and similar in case of smooth cycles) the eigenvector space will be similar to DFT.

In this paper we propose that  weight of an edge in the graph is a function of the variation between the signals at its incident vertices. Such graphs have been used before for example for image denoising in~\cite{CouGraNajPesTal} and edge aware image processing~\cite{NarYunOrt}.  We {\em reduce} this weighted graph to its minimal weight vertex-disjoint cycle cover (VCC). This is a reduction and not necessarily a decomposition as in~\cite{EkaFanAyaRam}, because a VCC may not contain all the edges of the graph. As the weights represent the variations in the signal we expect that VCC has cycles that are intrinsically smooth.   Some of the discarded edges maybe chords of a cycle, while others maybe between cycles, we contend that these edges have larger weights and hence connect vertices with large signal changes. Having found the cycles we can then  process signals on vertices of each cycle using traditional LSI filters -- thus all signals on a graph are processed using predominately low frequency dependencies.  In this paper we show that an image, modelled by a 8-lattice, can be reduced to a VCC, and use this for a simple image denoising algorithm.

Though we have motivated our work from the graph signal processing perspective, it has commonality, especially for lattice based image graphs, with mesh processing, e.g.,~\cite{KarGot},  discrete calculus based image processing, e.g.,~\cite{CouGraNajPesTal}, or with graph cuts based image segmentation, e.g.~\cite{JiaMal}. Like these works we partition the graph based on variation of signals on vertices of the graph, but our partitions are not arbitrary subgraphs, rather are cyclic subgraphs so that we can process the signals using LSI systems~\cite{EkaFanAyaRam}.  The relationship between graph signal processing and these traditional works has been commented upon in~\cite{ShuNarFroOrtVan}, it needs to be understood in greater depth. Note that we can use our algorithm on any graph not only a mesh or a lattice; we believe that the main advantages may lie for processing graphs with non-regular topology.  In section~\ref{sec-mm} we sketch the VCC algorithm, and in section~\ref{sec:disc} discuss the current results.

\section{Materials and Methods}
\label{sec-mm}
Following terminology introduced in~\cite{EkaFanAyaRam}, consider an undirected, simple, connected graph $G = (V,E)$, where $V$ is the set of $N$ vertices, and $E$ is the set of edges. Let $A$ be the adjacency matrix of the graph, where $A(i, j)$ is the nonnegative edge weight between nodes $i$ and $j$; $A(i, j )$ is zero if no edge connects the two nodes. A signal defined on the graph is a vector ${\bf{x} }: V \rightarrow$ $\mathbb{R}^N$, where $x(v)$ denotes the value of the signal at vertex $v$ in $V$.  Let $D$ be a diagonal matrix, where the diagonal entry $D(i, i) = \sum_{j=0}^{N-1} A(i, j)$. The Laplacian matrix of the graph is defined as $L = D - A$. It is a positive semi-definite matrix and has the spectral decomposition $L = U\Lambda U^H$, where $\Lambda$ is a diagonal matrix of non-negative real eigenvalues. The graph eigenvectors $\{\bf{u_k} \}^{N-1}_{k=0}$, constitute an orthonormal basis  for $R^N$. The corresponding eigenvalues for a connected graph are ${0=\lambda_0 <\lambda_1 \leq\lambda_2 \leq\hdots\leq\lambda_{N-1}}$. 

The book~\cite{LovPlu} [chapter 10] describes Tutte's reduction method that is used to find vertex-disjoint cycle cover of an undirected $G$, but for the sake of completion we will sketch it in algorithmic terms here. For this discussion $G$ is assumed to be unweighted. Let us start with some definitions: a {\em matching} $M$ in $G$ is a set of pairwise non-adjacent edges; that is, no two edges share a common vertex. A {\em perfect matching}  is a matching which matches all vertices of the graph, i.e., for every $v \in V$ there is an edge incident on $v$ in $M$. Assume that an integer $f(v)$ is assigned to each vertex in $G$, a {\em f-factor} of a graph $G$ is a spanning subgraph $H$ of $G$ such that $deg_H(v) = f(v)$. A {\em perfect f-matching} is to assign a non-negative integer $n(e)$ to every edge $e$ such that $\sum_{e = (*,v)} n(e) = f(v);\forall v$, where the notation $e=(*,v)$ implies all edges incident on $v$. 

Tutte's reduction of $G = (V,E)$ is based on construction of two graphs $H_G = (U, E_H)$ and $F_G = (V\bigcup V', E_F)$. To construct $H_G$: for each $v \in V$, let $U_v$ be a set of $f(v)$ vertices, such that $U  = \bigcup_{v \in V} U_v$ and $U_v  \bigcap U_w = \emptyset$ if $v\neq w$. For each edge $e = (v,w) \in G$, connect each vertex of $U_v$  to each vertex of $U_w$ in $H$.  On the other hand to construct $F_G$: let $V'=\emptyset, E_F = \emptyset$. For each edge $e = (v,w) \in E$, add two vertices $e_v, e_w$ to $V'$, and connect $v$ to $e_v$, $e_v$ to $e_w$, and $e_w$ to $w$ in $F$.

From above a 2-factor is equivalent to the vertex cycle cover of $G$.   Lov$\acute{a}$sz and Plummer have  shown that there exists a 2-factor in G {\em iff} there exists a perfect 2-matching in $F_G$. Further there exists a perfect 2-matching in $F_G$ {\em iff} there exists a perfect matching in  $H_{F_G}$.  Thus a vertex disjoint cycle cover  of $G$ is equivalent to a perfect matching of $H_{F_G}$~\cite{LovPlu}. The algorithm is described below:
\begin{itemize}
\item Remove  degree one vertices of $G$ recursively.
\item Construct $F_G = (V \bigcup V', E_F)$ as described above.
\item Let $T=(\bigcup_{v \in V} U_v) \bigcup (\bigcup_{v'\in V'} U_{v'})$. Construct $H_{F_G} = ( T, E_H)$ as described above, with $f(v) = 2\;\forall v \in V$ and $f(v) =1\;\forall v \in V'$.
\item Use Edmond's algorithm~\cite{Edm} to find perfect matching $M=(T,E_M)$ in $H_{F_G}$. We use the implementation by Kolmogrov~\cite{Kol} which can also find minimum weight perfect matching in case of weighted graphs.
\item Construct a graph $M' = (V \bigcup V', E_{M'})$ such that if there is an edge in $M$ between any element of $U_v$ and any element of $U_w$, there is an edge between $v$ and $w$ in $E_{M'}$, where $v,w \in V \bigcup V'$. 
\item Construct a graph $C = (V, E_C)$: for each edge $e \in E_{M'}$,
\begin{itemize}
\item If $e = (v, e_w)$, $v \in V$ {AND} $e_w \in V'$, 
\item Or $e = (e_v,w) $, $e_v \in V'$ {AND} $w \in V$,
\item Or $e = (v,w)$,  $v \in V$ {AND} $w \in V$
\begin{itemize}\item add an edge $(v,w)$ to $E_C$.
\end{itemize}
\end{itemize}
\item Graph $C$ is the vertex-disjoint cycle cover of $G$.
\end{itemize}

Note that if $G$ is weighted the degree of a vertex is defined by the number of edges incident on it, not by the sum of the weight of the edges. Also when adding edges to $H_G$ or $F_G$ the weight of the corresponding edge in $G$ is used as the weight of the resultant edge. The complexity of the Edmond's algorithm for a $G=(V,E)$ is $O(\sqrt{|V|}|E|)$.  In $H_{F_G}$ there are $2*|V|+2*|E|$ vertices and  $9*|E|$ edges for a given graph $G=(V,E)$. Thus the complexity of perfect matching in $H_{F_G}$ is $O(\sqrt{|V|+|E|}|E|)$. The complexity of construction $H_{F_G}$ and $C$ is relatively small and negligible respectively. A paper by Manthey~\cite{Man} further discusses the complexity of finding minimum weight cycle covers in graphs and its approximability.

The existence of perfect matching, and hence vertex-disjoint cycle cover, in a graph is dictated by the Tutte theorem~\cite{LovPlu}. For  8-lattice graphs modelling images we did not come across situations where the cover did not exist. A simple solution to the case where a VCC does not exist would be to add random edges with very high weight to the graph. These edges will be part of the solution if and only if no other cycle with lower total weight exists. In the resultant these random edges can be removed, and hence the cover may contain open cycles. This idea and others have to be explored in greater detail.
\begin{figure}[!t]
  \centering
  \centerline{\includegraphics[width=9.2cm,height=8cm]{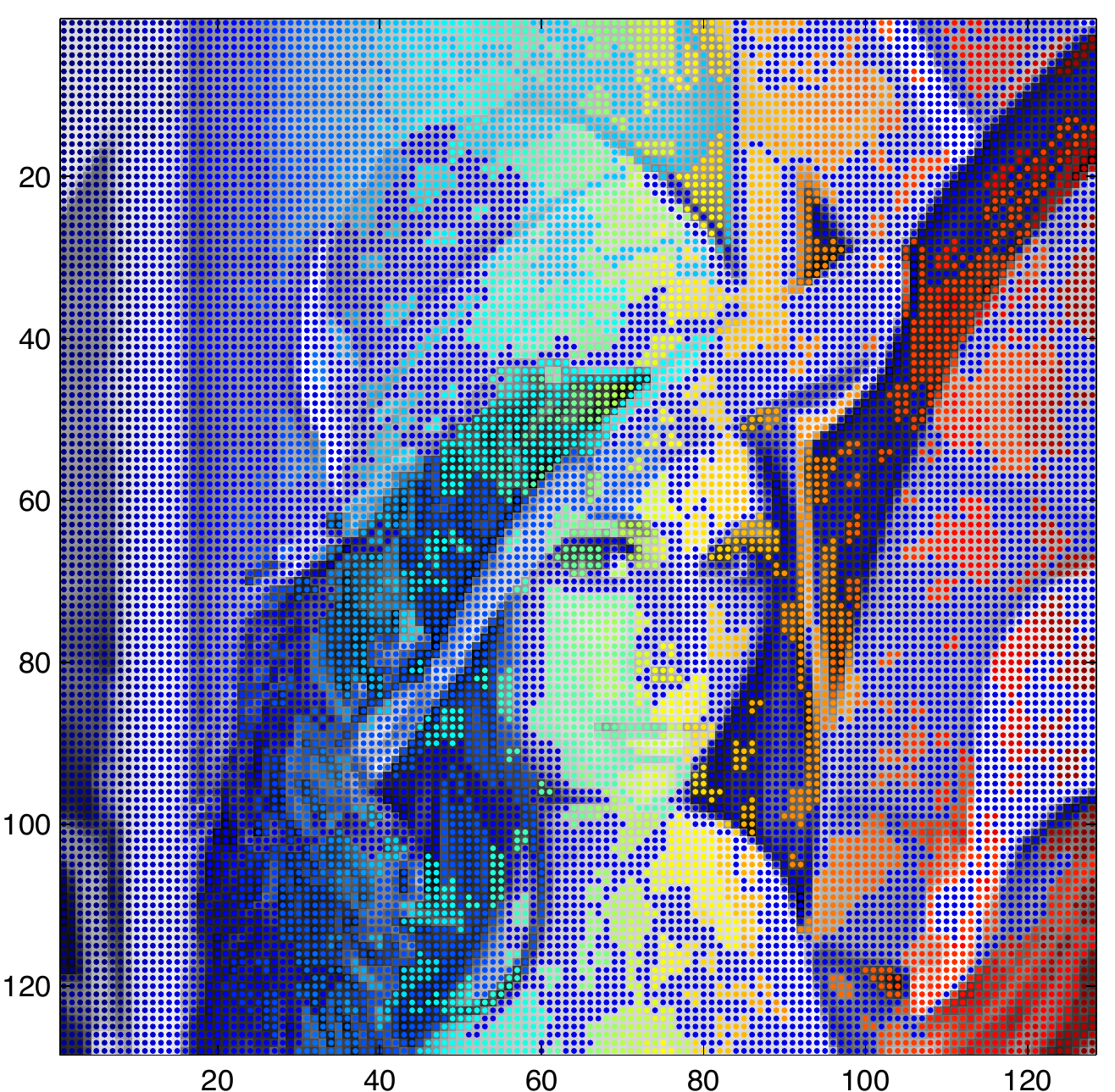}}
 \caption{VCC of Lena image. Pixels belonging to the same cycle are marked with the same colour. Different cycles are marked with different colours uniformly sampled from the "Jet" colormap of Matlab. Edge weights of the graph are exponential of the difference between pixel values of the vertices incident on the edge~\cite{GraphToolBox}. Weights are quantised to an integer using their index in ascending sorted order.  }
\label{fig-resLenna1}
\end{figure}
\begin{figure}[]
  \centering
  \centerline{\includegraphics[width=9.2cm,height=8cm]{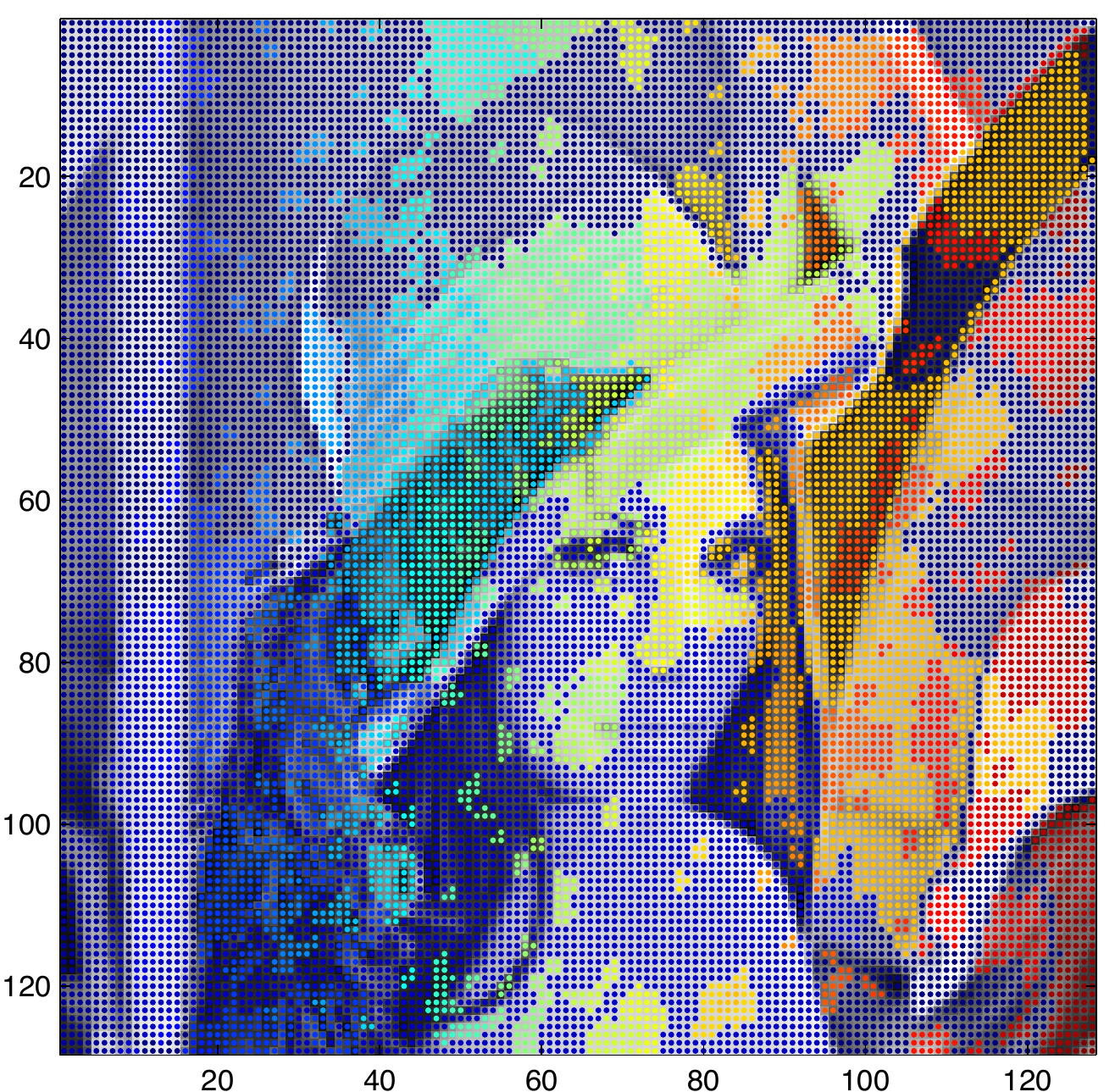}}
 \caption{VCC of Lena image. 
Here edge weights of the graph are assigned using an {\em edge map} extracted using the Canny {\em edge} detector. Edge weights are binary: for {\em edge} to {\em non-edge} pixels in the image the weights are high (integer value 5), else they are low (integer value 1)~\cite{NarYunOrt}.   }
\label{fig-resLenna2}
\end{figure}

\section{Discussion}
\label{sec:disc}
\begin{figure}
  \centering
  \centerline{\includegraphics[width=9.2cm,height=8cm]{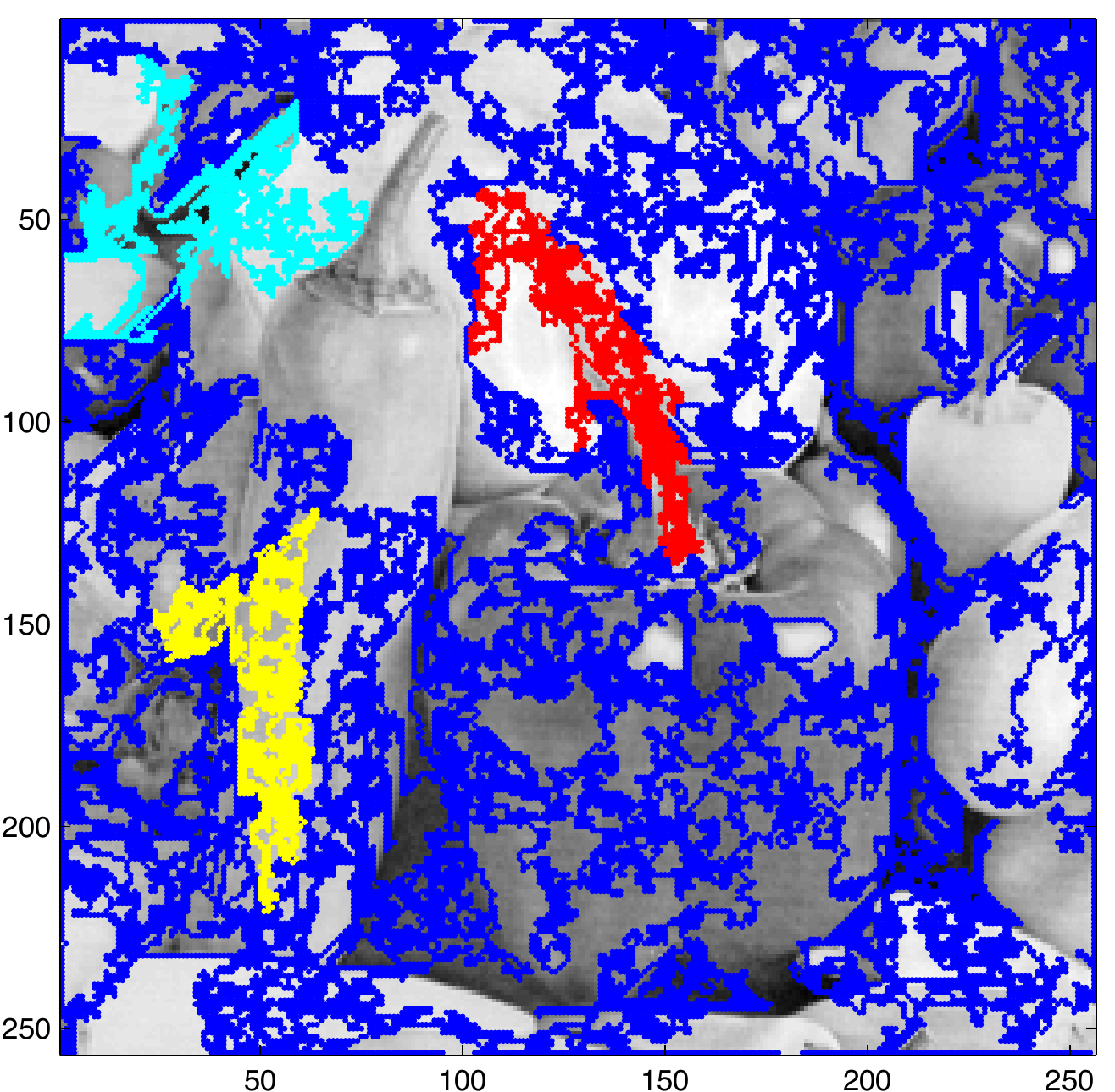}}
\caption{VCC of the Peppers image. Only the largest 4 cycles are shown for clarity. Edge weights of the graph are exponential of the difference between pixel values of the vertices incident on the edge.  }
\label{fig-resPep}
\end{figure}

Images are modelled by a 8-lattice, we use the graph tool box by Grady and Schwatrz~\cite{GraphToolBox} to construct the image graph. Fig.~\ref{fig-resLenna1}, \ref{fig-resLenna2} show the VCC for Lena image for two different edge weights. 
In Fig.~\ref{fig-resPep} the four largest cycles of the VCC of the peppers image are shown for clarity.  From these figures one can see that though the cycles tend to follow contours, they do {\em not} segment the image into background/foreground, or different objects.

\begin{figure}[!b]
  \centering
  \centerline{\includegraphics[width=9.2cm,height=5.0cm]{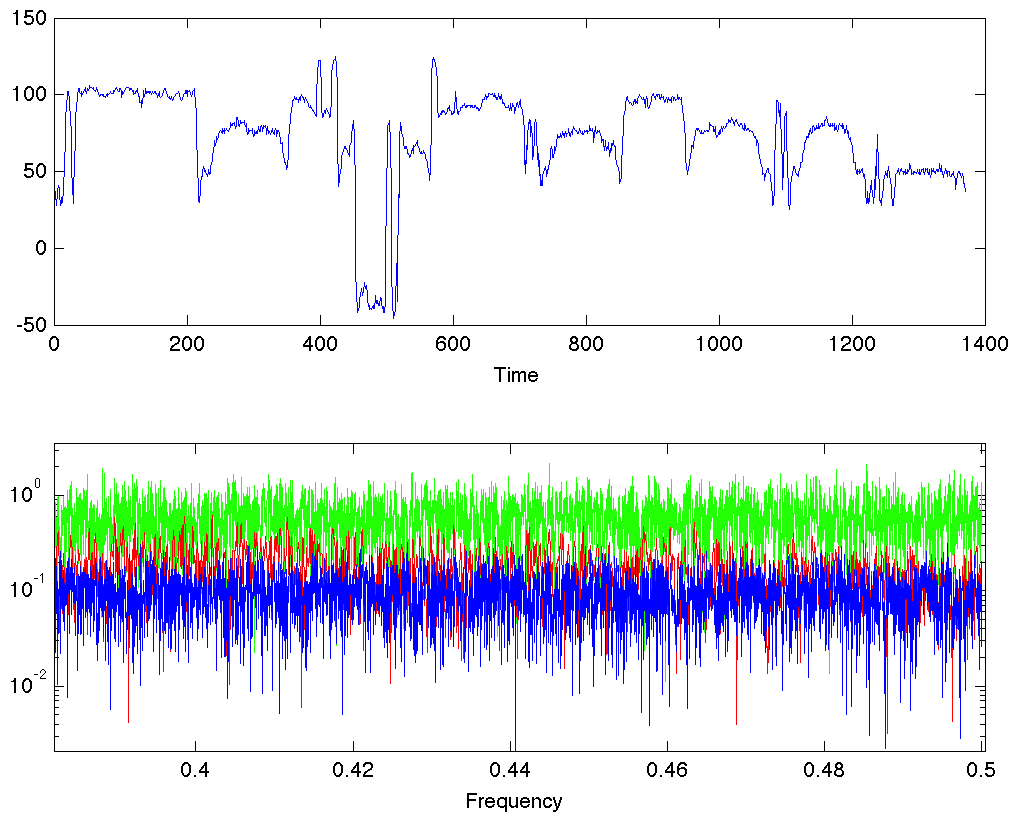}}
 \caption{ For Lena image, pixel values along a randomly chosen cycle are shown in the top plot. In the bottom plot the higher end frequency spectrum of three cases is shown in semilogx scale: Blue is the spectrum of the longest VCC cycle in Fig.~\ref{fig-resLenna1}. Green is the  spectrum of this cycle after random permutation. Red is the  spectrum of the longest cycle in VCC of an image graph where all edge weights are set to 1.}
\label{fig-resSigFreq}
\end{figure}
When difference in pixel value is used as the edge weight, Fig.~\ref{fig-resLenna1}, the mean variation along edges in the image graph is 17.60. The mean variation along edges in the VCC of this graph is 5.15, and the mean variation along edges not in VCC is 21.87. VCC has approximately 25\% of the edges, but only 7\% of the total edge weight. 
 A random cycle is chosen from the VCC, Fig.~\ref{fig-resLenna1}, and its pixel values are plotted in the top plot of Fig.~\ref{fig-resSigFreq}. The plot shows a step-like signal which is   smooth in intervals. The bottom plot of the same figure shows that the high frequency spectrum of the largest cycle in this VCC has lower energy than a randomly permuted cycle, or a cycle of a graph where all the edge weights are one.  These results illustrate that the proposed method finds smooth cycles.

%
\begin{figure}[]
  \centering
  \centerline{\includegraphics[width=9.2cm,height=8cm]{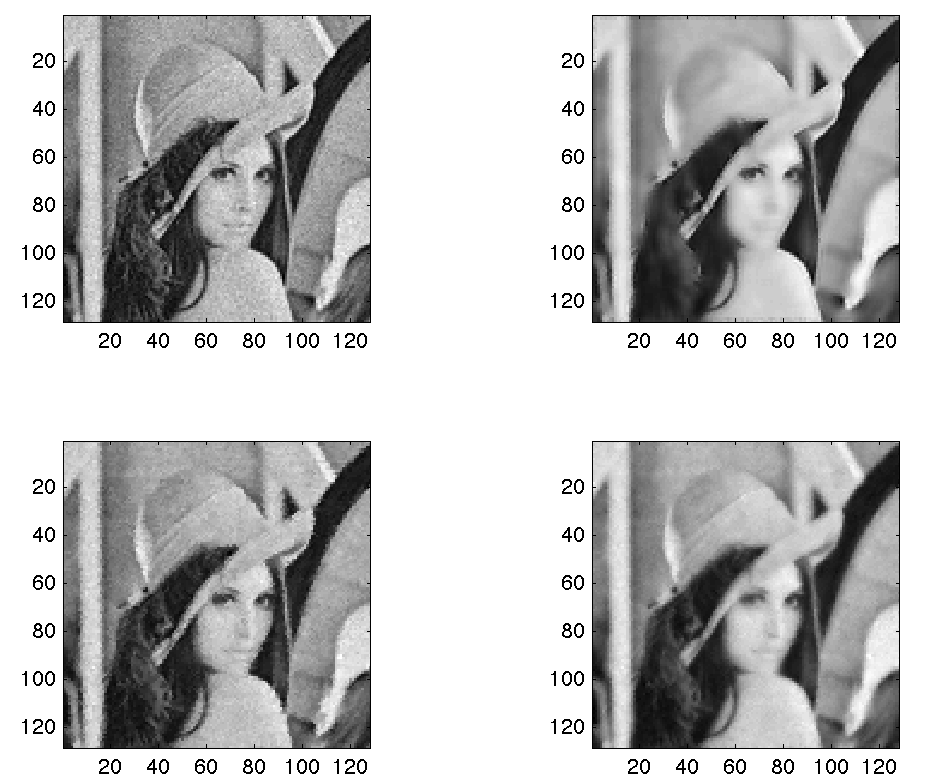}}
 \caption{Top left: Noisy Lena image. Top Right: Denoised using Wiener filter. Bottom Left: Denoised using VCC+GFT. Bottom Right: Denoised using GFT.}
 \label{fig-resNoise}
\end{figure}
As an application we have preliminary results from image denoising using the Tikhnov regularization as explained in example two of~\cite{ShuNarFroOrtVan} .  Noise is generated using Gaussian process, and for result shown here, Fig.~\ref{fig-resNoise}, with mean zero and standard deviation 7. In ``GFT" we take the graph Fourier transform of the entire image and  regularise  frequency components, while in ``VCC+GFT" we take the Fourier transform of pixels in a cycle (in their cyclic) order and regularise frequency components of each cycle independently. ``VCC+GFT" has the sharpest image in the result, however it introduces artefacts because neighbours in the lattice may be processed by  different cycles.

\begin{figure}[]
\vspace{-1cm}

  \centering
  \centerline{\includegraphics[width=9.2cm, height=4cm]{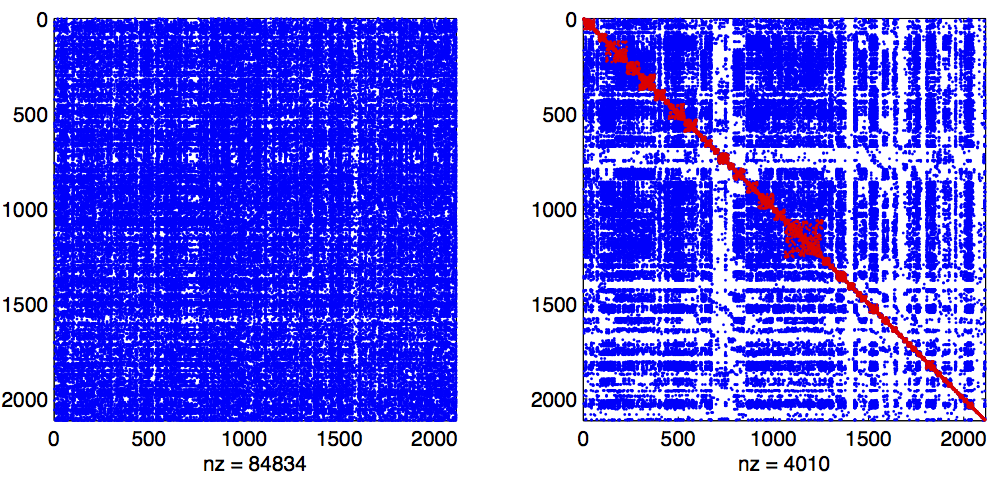}}
 \caption{Left plot is the adjacency matrix of a graph where the vertices are selected images from Flickr, and the edges encode a measure of images' similarity in the colour space~\cite{SinICASSP2013}. Right plot, the VCC of this graph is plotted using red dots. Further the original adjacency matrix is reordered using the cycles of VCC and plotted using blue dots.}
\label{fig-resFlickr}
\end{figure}

We have also experimented with a graph with arbitrary topology, Fig.~\ref{fig-resFlickr} . VCC of the graph does not exist, adding random edges with very high weight to this graph, we have found an approximate VCC, which has both closed and open cycles. Surprisingly the overall composition of VCC is stable to multiple iterations of adding random edges. The figure shows that VCC is able to reorder the original adjacency matrix such that its entropy reduces, implying that its able to detect useful clusters, associations~\cite{ChaPapModFal}.

VCC show a promise for graph based signal processing. Its a simple idea -- that a graph be reduced to cycles and then signals on each cycle be independently processed using LTI theory. This is a exploratory work, we are working towards understanding it in greater detail.


\clearpage

\bibliographystyle{IEEEbib}
\bibliography{refs,/Users/raghavsi/Documents/Papers/mybib}

\end{document}